\documentclass[9pt,twocolumn,twoside]{pnas-new}
\templatetype{pnasresearcharticle} % Choose template 
\setboolean{displaywatermark}{false}
\usepackage{graphicx}% Include figure files
\usepackage{dcolumn}% Align table columns on decimal point
\usepackage{bm, amssymb, color}% bold math
\usepackage{hyperref}% add hypertext capabilities
\usepackage{natbib}
\usepackage{booktabs,tabulary}
\usepackage{multirow}
\usepackage{amsmath,soul}
\usepackage[export]{adjustbox}
\usepackage[caption=false]{subfig}
\usepackage{array}
\usepackage{cases}
\usepackage{makecell}
\usepackage{physics}
\usepackage{enumitem}

\newcommand{\R}{\mathbb{R}}

\newcommand{\fn}[2]{\mathinner{#1\mathopen{\left(#2\right)}}}
\newcommand{\vect}[1]{\bm{#1}}

\newcommand{\spD}[1]{\fn{\tilde{\chi}_{_V}}{#1}}
\newcommand{\tens}[1]{\mathinner{\mathbf{#1}}}

\newcommand{\cL}{\mathinner{c_{{_L}_1}}}
\newcommand{\cT}{\mathinner{c_{{_T}_1}}}
\newcommand{\kL}{\mathinner{k_{{_L}_1}}}
\newcommand{\kT}{\mathinner{k_{{_T}_1}}}

\newcommand{\cLp}{\mathinner{c_{{_L}_p}}}
\newcommand{\cTp}{\mathinner{c_{{_T}_p}}}

\newcommand{\Hankel}[2]{\fn{H_{#1}^{(1)}}{#2}}
\newcommand{\uvect}[1]{\hat{\vect{#1}}}

\usepackage[colorinlistoftodos,shadow,textwidth=18mm]{todonotes}

\setstcolor{red}

\setul{0}{0.3ex}

\title{
Multifunctional Composites for Elastic and Electromagnetic Wave Propagation}

\author[a]{Jaeuk Kim}
\author[a,b,c,d]{Salvatore Torquato} 

\affil[a]{Department of Physics, Princeton University, Princeton, New Jersey 08544, USA}
\affil[b]{Department of Chemistry, Princeton University, Princeton, New Jersey 08544, USA}
\affil[c]{Princeton Institute for the Science and Technology of Materials, Princeton University, Princeton, New Jersey 08544, USA}
\affil[d]{Program in Applied and Computational Mathematics, Princeton University, Princeton, New Jersey 08544, USA}

\leadauthor{Jaeuk Kim} 

\significancestatement{
We establish accurate microstructure-dependent cross-property relations for composite materials that link effective elastic and electromagnetic wave characteristics to one another, including effective wave speeds and attenuation coefficients.
    Our microstructure-dependent formulas enable us to explore the multifunctional wave characteristics of a broad class of disordered microstructures, including exotic disordered ``hyperuniform'' varieties, that can have advantages over crystalline ones, such as nearly optimal, direction-independent properties and robustness against defects. 
    Applications include filters that transmit or absorb elastic or electromagnetic waves ``isotropically'' for a range of wavelengths.
    Our findings enable one to design multifunctional composites via inverse techniques, including exterior components of spacecraft or building materials, heat-sinks for CPUs, sound-absorbing housings for motors, and nondestructive evaluation of materials.
     }

\authorcontributions{Author contributions: S.T. designed research; J.K and S.T. performed research, analyzed data, and wrote the paper.
}
\authordeclaration{The authors declare no conflict of interest}
\equalauthors{\textsuperscript{1}J.K. and S.T. contributed equally to this work.}
\correspondingauthor{\textsuperscript{2}To whom correspondence should be addressed. E-mail: \href{mailto:torquato@princeton.edu}{torquato@princeton.edu}}


\begin{abstract}

	Composites are ideally suited to achieve desirable multifunctional effective properties since the best properties of different materials can be judiciously combined with designed microstructures. 
	Here we establish cross-property relations for two-phase composite media that link effective elastic and electromagnetic wave characteristics to one another, including the respective effective wave speeds and attenuation coefficients, which facilitate multifunctional material design.
    This is achieved by deriving accurate formulas for the effective electromagnetic and elastodynamic properties that depend on the wavelengths of the incident waves and the microstructure via the spectral density.
    Our formulas enable us to explore the wave characteristics of a broad class of disordered microstructures because they apply, unlike conventional formulas, for a wide range of incident wavelengths, i.e., well beyond the long-wavelength regime.
    This capability enables us to study the dynamic properties of exotic disordered ``hyperuniform'' composites that can have advantages over crystalline ones, such as nearly optimal, direction-independent properties and robustness against defects.
	We specifically show that disordered ``stealthy'' hyperuniform microstructures exhibit novel wave characteristics, e.g., low-pass filters that transmit waves ``isotropically'' up to a finite wavenumber.
    Our cross-property relations for the effective wave characteristics can be applied to design multifunctional composites via inverse techniques.
    Design examples include structural components that require high stiffness and electromagnetic absorption, heat-sinks for CPUs and sound-absorbing housings for motors that have to efficiently emit thermal radiation and suppress mechanical vibrations, and nondestructive evaluation of the elastic moduli of materials from the effective dielectric response.

\end{abstract}

\dates{This manuscript was compiled on \today}
\doi{\url{www.pnas.org/cgi/doi/10.1073/pnas.XXXXXXXXXX}}

\begin{document}

\maketitle
\thispagestyle{firststyle}
\ifthenelse{\boolean{shortarticle}}{\ifthenelse{\boolean{singlecolumn}}{\abscontentformatted}{\abscontent}}{}

\dropcap{A} heterogeneous material (medium) consists of domains of multiple distinct materials (\textit{phases}).
Such materials are ubiquitous; examples include sedimentary rocks, particulate composites, colloidal suspensions, polymer blends, and concrete \cite{ Torquato_RHM,Milton_TheoComposites, Zohdi2012, Neville_concrete, Sahimi_HM1, Turner_fabrication_2011}. 
When domain (inhomogeneity) length scales $\ell$ are much smaller than the system size, a heterogeneous material can be regarded as a homogeneous material with certain effective physical properties, such as thermal (electric) conductivity $\sigma_e$, dielectric tensor $\tens{\epsilon}_e$, or stiffness tensor $\tens{C}_e$ \cite{Sahimi_HM1, Milton_TheoComposites, Torquato_RHM}.
Such effective properties depend on the phase properties, phase volume fractions $\phi_i$, and higher-order microstructural information \cite{Beran1968, Dederichs1973, Torquato1985, Sen1989,  Torquato1997, Rechtsman2008, Torquato_RHM, Sahimi_HM1, Milton_TheoComposites}.
	Heterogeneous materials are ideally suited to achieve \textit{multifunctionality}, since the best features of different materials can be combined to form a new material that has a broad spectrum of desired properties  \cite{Torquato1990, Gibiansky1993, Torquato2002, Torquato_RHM, Torquato2018_5, Milton1981b, Luis2007, Wang2016}. 
	Because the effective properties of a heterogeneous material reflect common morphological information, knowledge of one effective property can provide useful information about a different effective property \cite{Torquato1990, Gibiansky1993, Torquato2002, Torquato_RHM, Torquato2018_5, Milton1981b, Luis2007, Wang2016}.
	Such \textit{cross-property} relations can aid in the rational design of multifunctional heterogeneous materials that possess multiple desirable effective properties \cite{Torquato1990, Gibiansky1993, Torquato2002, Torquato_RHM, Torquato2018_5, Milton1981b, Luis2007, Wang2016} via {\it inverse} techniques \cite{Torquato2009_inverse}.

All of the previous applications of cross-property relations for multifunctional design have focused on the static transport and elastic properties.
Remarkably, however, nothing is known about analogous cross-property relations for various effective dynamic properties, each of which is of great interest in its own right.
	For example, in the case of propagation of electromagnetic waves in two-phase media, the key properties of interest is the frequency-dependent dielectric constant, which is essential for a wide range of applications, including remote sensing of terrain \cite{tsang_theory_1985}, investigation of the microstructures of biological tissues \cite{sihvola_electromagnetic_1999}, probing artificial materials \cite{zhuck_strong-fluctuation_1994}, studying wave propagation through turbulent atmospheres \cite{tatarskii_effects_1971}, investigation of  electrostatic resonances \cite{mcphedran_electrostatic_1980}, and 
	design of materials with desired optical properties \cite{sihvola_electromagnetic_1999, Wu2017}.	
	An equally important dynamic situation occurs when elastic waves propagate through a heterogeneous medium, which is of great importance in geophysics \cite{Biot1956, Guy1974}, exploration seismology \cite{Sheriff1995}, diagnostic sonography \cite{Sarvazyan2013}, crack diagnosis \cite{Sutin1995}, architectural acoustics \cite{Watson_RoomAcoustics} and acoustic metamaterials \cite{Yuan_2019_energy-harvesting}.

	\begin{figure}[h!]
  \begin{center}
\subfloat[]{\includegraphics[width=0.49\textwidth]{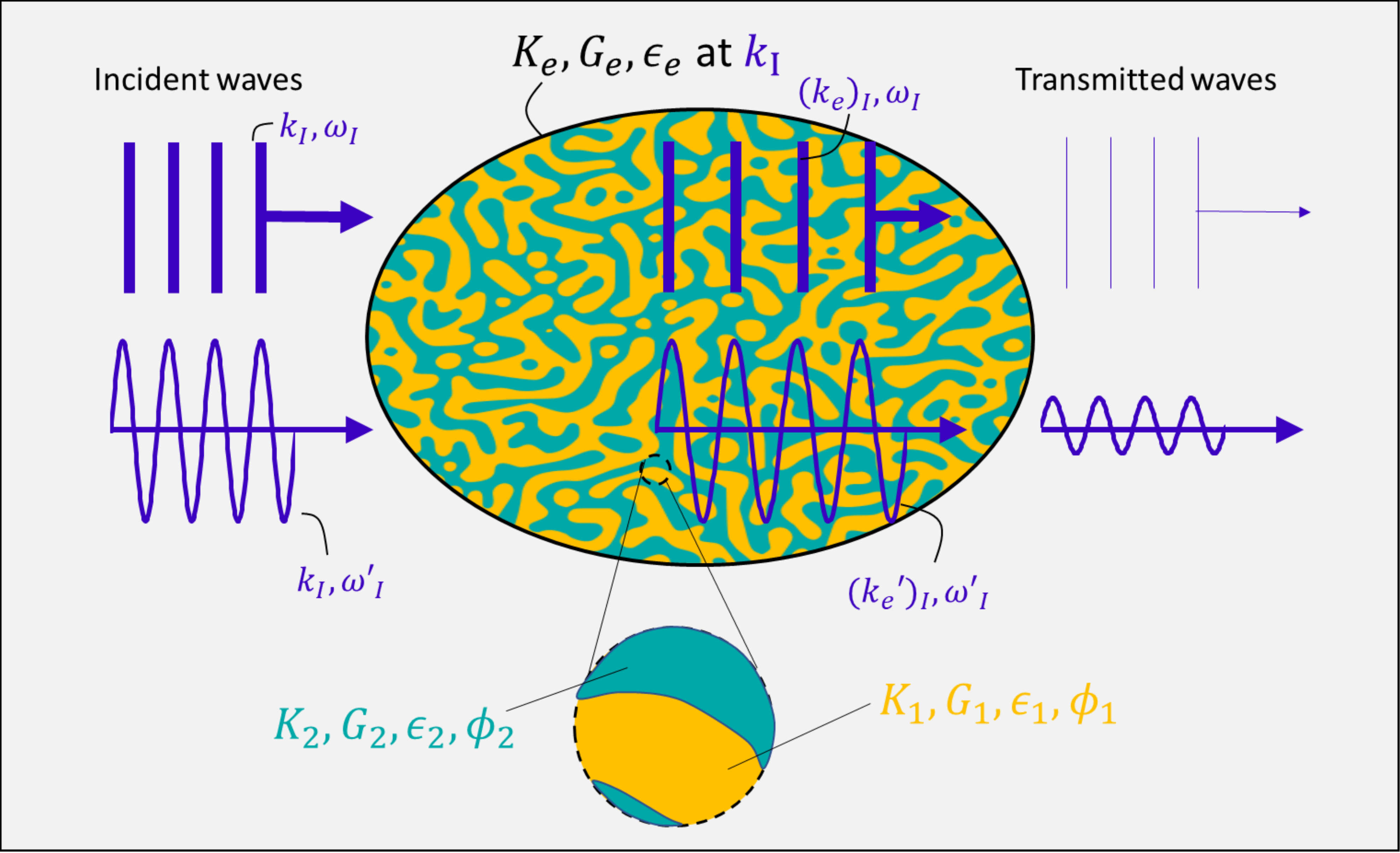}}
\hspace{5pt}
\subfloat[]{\includegraphics[width=0.49\textwidth]{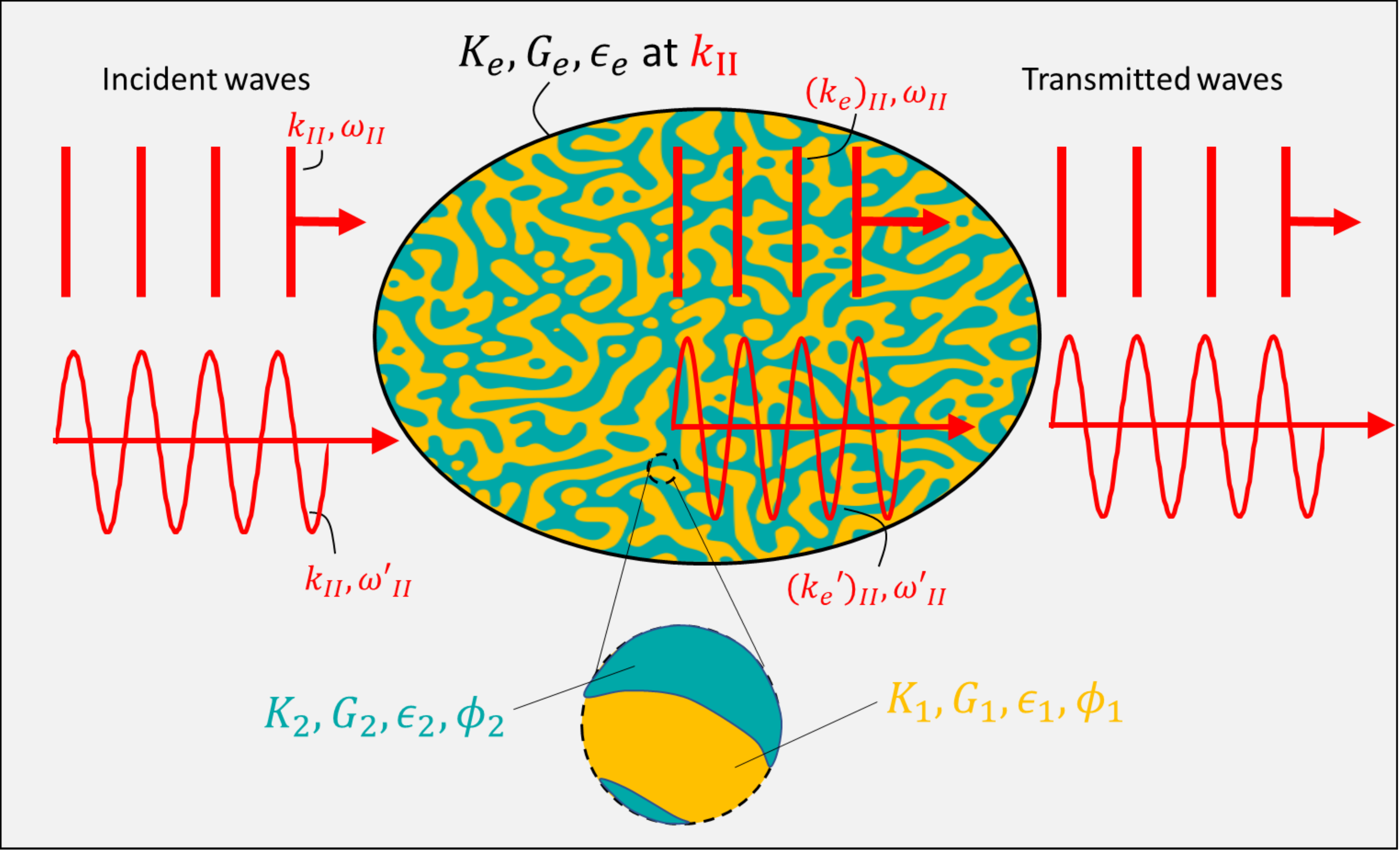}}
  \end{center}
\caption{Schematics illustrating multifunctional applications of heterogeneous materials. 
Elastic and electromagnetic waves at two different wavenumbers (a) $k_I$ and (b) $k_{II}$  incident to, inside of and transmitted from a composite material (a large ellipse)  consisting of a matrix phase (shown in yellow) and a dispersed phase (shown in cyan).
Parallel lines and sinusoidal curves represent elastic and electromagnetic waves, respectively. 
(a) For an elastic wave  with a wavenumber $k_I$, while the wavefronts inside this material  experience microscopic disturbances, they effectively behave like a plane wave inside a homogeneous material with an effective wavenumber $(k_e)_{I} = \omega_I [c_e^L + i \gamma_e^L]^{-1}$ and effective elastic moduli $K_e$ and $G_e$; see definition \ref{eq:effective_wave_propagations}. 
 Analogously,  for an electromagnetic wave, this material behaves like a homogeneous material with an effective dielectric constant $\epsilon_e$.
For instance, both elastic and electromagnetic waves are attenuated due to scattering if this  composite has a non-zero scattering intensity at $k_I$ [i.e., $\gamma_e^L < 0$ and $\Im[\epsilon_e] > 0$]. 
 (b) For both types of waves (red) of a wavenumber $k_{II}$, this composite can be effectively transparent, if it has a zero-scattering intensity at $k_{II}$ [i.e., $\gamma_e^L = 0$ and $\Im[\epsilon_e]=0$].
\label{fig:schem}}
\end{figure}

	Our study is motivated by the increasing demand for multifunctional composites with desirable wave characteristics for a specific bandwidth (i.e., a range of frequencies).
	Possible applications include sensors that detect changes in moisture content and water temperature \cite{ekmekci_multi-functional_2013}, thin and flexible antennas \cite{ali_design_2019}, materials that efficiently convert acoustic waves into electrical energy \cite{mikoshiba_energy_2013}, materials that can attenuate low-frequency sound waves and exhibit excellent mechanical strength \cite{tang_hybrid_2017}, and materials with negative modulus in the presence of magnetic fields \cite{yu_magnetoactive_2018}; see Ref. \cite{lincoln_multifunctional_2019} and references therein.

	However, systematic design of multifunctional materials with desirable elastodynamic and electromagnetic properties has yet to be established.
	In this paper, we derive accurate microstructure-dependent formulas for the effective dynamic dielectric constant $\epsilon_e$ and effective dynamic bulk $K_e$ and shear $G_e$ moduli, each of which depends on the appropriate wavelength (or wavenumber) of the incident waves. 
	We show that these formulas can accurately capture the dynamic responses of composites that are valid for a wide range of wavelengths, i.e., beyond the long-wavelength limitation of conventional approximations. 
	From these effective properties, one can obtain the effective wave speed $c_e$ and attenuation coefficient $\gamma_e$ for the electromagnetic waves, and the analogous quantities $c_e^{L,T}$ and $\gamma_e^{L,T}$ for the longitudinal (L) and transverse (T) elastic waves.
	From these formulas, we are able to derive cross-property relations that link electromagnetic and elastodynamic properties to one another.
	Such cross-property relations facilitate multifunctional design.
	Two striking multifunctional design applications are schematically illustrated in Fig. \ref{fig:schem}.

	The challenge in deriving cross-property relations is that the effective properties depend on an infinite set of correlation functions \cite{Sahimi_HM1, Milton_TheoComposites, Brown1955, Beran1968, Dederichs1973, Torquato1985, Milton1987a, Milton1987b, Sen1989,  Torquato1997, Torquato_RHM, Rechtsman2008}.
	To derive the pertinent cross-property relations for the aforementioned effective wave characteristics, we rely on {\it strong-contrast expansions} \cite{Brown1955, Torquato1985, Sen1989, Torquato1997, Torquato_RHM, Rechtsman2008}.
	The strong-contrast formalism represents a very powerful theoretical approach to predict the effective properties of composites for any phase contrast and volume fraction \cite{Brown1955, Torquato1985, Sen1989, Torquato1997, Torquato_RHM, Rechtsman2008}.
	They are formally exact expansions whose terms involve functionals of the $n$-point correlation function $\fn{S^{(i)}_n}{\vect{x}_1, \cdots,\vect{x}_n}$ for all $n$ and field quantities as well as a judicious choice of the expansion parameter \cite{Torquato1985, Sen1989, Torquato1997, Torquato_RHM, Rechtsman2008}.
	Here, the quantity $\fn{S_n ^{(i)}}{\vect{x}_1, \cdots, \vect{x}_n}$ gives the probability of finding $n$ points at positions $\vect{x}_1, \cdots, \vect{x}_n$ simultaneously in phase $i$.
	Remarkably, the rapid convergence of strong-contrast expansions has enabled one to extract accurate estimates of the effective properties of a wide class of composites (dispersions of particles in a matrix) by approximately accounting for complete microstructural information.
	Specifically, higher-order functionals are approximated in terms of lower-order functionals; see Sec. VI.B in the {\it\color{blue}SI}. 
	Such microstructure-dependent approximations have been obtained for the effective static dielectric constant \cite{Torquato1985, Sen1989}, effective static stiffness tensor \cite{Torquato1997}, and the effective dynamic dielectric constant \cite{Rechtsman2008}.

	In the latter instance involving electromagnetic waves, the wavenumber-dependent effective dielectric constant $\fn{\epsilon_e}{k_1}$ for macroscopically isotropic two-phase composites depends on a functional involving the {\it spectral density} $\spD{\vect{Q}}$ \cite{Rechtsman2008}. 
	The quantity $\spD{\vect{Q}}$ is the Fourier transform of the \textit{autocovariance} function $\fn{\chi_{_V}}{\vect{r}} \equiv \fn{S_2 ^{(i)}}{\vect{r}} - {\phi_i}^2$, where $\vect{r}\equiv\vect{x}_2 - \vect{x}_1$, and can be measured from scattering experiments \cite{Debye1949}.
	In principle, this approximation is valid only in the long-wavelength regime, i.e., $k_1 \ell \ll 1$, where $k_1$ is the wavenumber of the electromagnetic waves in the reference phase (phase 1).
	However, we modify this formula in order to extend it to provide a better estimate down to the intermediate-wavelength regime (i.e., $k_1\ell \lesssim 1$) by accounting for spatial correlations of the incident plane waves; see Eq. \ref{eq:strong-contrast_dielectric_2pt} in {\bf Results}. 
	This modified formula is superior to the commonly employed Maxwell-Garnett approximation \cite{sihvola_electromagnetic_1999, markel_maxwell_2016} that, unlike formula \ref{eq:strong-contrast_dielectric_2pt}, fails to capture salient physics in correlated disordered systems; see Sec. V in the {\it\color{blue}SI}.
A capacity to accurately predict the effective dielectric constant is essential for the aforementioned applications \cite{tsang_theory_1985, zhuck_strong-fluctuation_1994, tatarskii_effects_1971, mcphedran_electrostatic_1980, sihvola_electromagnetic_1999, Wu2017}.

To obtain analogous microstructure-dependent formulas for the effective dynamic elastic moduli $\fn{K_e}{\kL}$ and $\fn{G_e}{\kL}$, we utilize strong-contrast expansions for them that we have derived elsewhere that also apply in the long-wavelength regime.
Here, $\kL$ is the wavenumber of longitudinal elastic waves. 
This dynamic formulation is considerably much more challenging mathematically than its dielectric counterpart \cite{Rechtsman2008} because of the complexity and nature of the fourth-rank tensor Green's functions that are involved.
In the present work, one primary objective is to extract from these expansions (see {\it\color{blue} Methods} for the formal expansion) accurate approximate formulas that also depend on the spectral density $\spD{\vect{Q}}$. 
As we did in the case of the dielectric formula, we modify these strong-contrast approximations for the effective dynamic moduli so that they are valid at the extended wavelengths ($\kL \ell \lesssim 1$).
In {\bf Results}, we employ these modified formulas to investigate the effective elastic wave characteristics, including effective wave speeds $c_e^{L,T}$ and attenuation coefficients $\gamma_e^{L,T}$, for four models of disordered dispersions. 
Knowledge of the effective elastodynamic properties is of importance in the aforementioned disciplines and applications; see Refs. \cite{Biot1956, Guy1974, Sheriff1995, Sarvazyan2013, Sutin1995, Watson_RoomAcoustics, Yuan_2019_energy-harvesting}.

We establish accurate cross-property relations linking the effective elastic and electromagnetic wave characteristics by utilizing the aforementioned microstructure-dependent formulas and by eliminating the common microstructural parameter among them. 
Thus, these results enable one to determine the response of a composite to electromagnetic waves from the corresponding response to acoustic/elastic waves and vice versa. 
The resulting cross-property relations will have practical implications, as discussed in the section {\bf Sound-absorbing and light-transparent materials} and {\bf Conclusions and Discussion}.

The primary applications we have in mind are disordered microstructures, both exotic and  ``garden'' varieties because they can provide advantages over periodic ones with high crystallographic symmetries, which include perfect isotropy and robustness against defects.
	Such disordered media have recently been exploited for applications involving photonic bandgap materials \cite{Florescu2009, Man2013}, gradient-index photonic metamaterials \cite{Wu2017}, compact spectrometers \cite{Redding2013}, random lasers \cite{Wiersma2013, Degl2016}, bone replacement \cite{Rabiei2009, Orivnak2014}, and impact-absorbers \cite{Garcia-Avila2014_armor, Marx2019}.

We are particularly interested in studying the wave characteristics of exotic disordered two-phase media, such as disordered \textit{hyperuniform} and/or \textit{stealthy} ones, and their potential applications. 
    Hyperuniform two-phase systems are characterized by anomalously suppressed volume-fraction fluctuations at long wavelengths \cite{Torquato2003_hyper, Zachary2009a, Torquato2018_review} such that 
\begin{equation}\label{eq:HU_condition}
\lim_{Q\to 0 }\spD{\vect{Q}} = 0,
\end{equation}
where $Q\equiv \abs{\vect{Q}}$ refers to a wavenumber.
    Such two-phase media encompass all periodic systems as well as exotic disordered ones; see Ref. \cite{Torquato2018_review} and references therein.
    These exotic disordered structures lie between liquids and crystals; they are like liquids in that they are statistically isotropic without any Bragg peaks, and yet behave like crystals in the manner in which they suppress the large-scale density fluctuations \cite{Torquato2003_hyper, Zachary2009a, Torquato2018_review}. 
    One special class of such structures are disordered \textit{stealthy hyperuniform} media that are defined by zero-scattering intensity for a set of wavevectors around the origin \cite{Uche2004, Torquato2015_stealthy, Zhang2015, Chen2017}.
	Such materials are endowed with novel physical properties \cite{Leseur2016, Zhang2016, Degl2016, Gkantzounis2017_freeform,Chen2017, Wu2017, Bigourdan2019}, including that they are transparent (dissipationless) to electromagnetic waves down to a finite wavelength \cite{Rechtsman2008, Leseur2016}.
    We also explore the wave characteristics of disordered \textit{stealthy} nonhyperuniform media that possess zero-scattering intensity for a set of wavevectors that do not include the origin \cite{Batten2008}.

	In the {\bf Conclusions and Discussion}, we describe how our microstructure-dependent estimates enable one to design materials that have the targeted attenuation coefficients $\gamma_e^{L,T}$ for a range of wavenumbers (or, equivalently, frequencies) via inverse-problem approaches \cite{Torquato2009_inverse}.
	Using the stealthy hyperuniform materials, we explicitly demonstrate that such engineered materials can serve as filters for elastic waves which selectively absorb \cite{Ma2013_acoustic, Chen2017_acoustic} or transmit \cite{Khelif2003, Romero2019} waves ``isotropically'' for a prescribed range of wavenumbers. 
	Furthermore, we show that we can engineer composites that exhibit {\it anomalous dispersion} \cite{Jackson_TEXTBOOK_3rd}, yielding resonance-like attenuation in $\gamma_e^{L,T}$.

\section*{Preliminaries} 
	We consider two-phase heterogeneous materials in $d$-dimensional Euclidean space $\R^d$.
For simplicity, the results reported here mainly focus on the case of $d=3$.
We also make three assumptions on the phase dielectric properties \cite{Rechtsman2008}: (a) the dielectric tensors of both phases are isotropic, (b) their dielectric constants are real-valued and independent of frequency, and (c) their magnetic permeabilities are identical. 

The three analogous assumptions for the elastodynamic problem are (a) both phases are elastically isotropic, (b) their elastic moduli are real numbers independent of frequency, and (c) they have identical mass densities ($\rho_1 = \rho_2$).
The last assumption is achievable for many pairs of solid materials, e.g., nickel, copper, and cobalt have mass densities about 8.9 $\mathrm{g/cm^3}$, and tin and manganese have mass densities about 7.2 $\mathrm{g/cm^3}$ \cite{Ashcroft}, but they have considerably different elastic moduli.

When these assumptions are met, inside each domain of phase $p~(=1,2)$, the elastic wave equation is given as \cite{Landau_elasticity}
$$
\omega^2 u_i +  \left({\cLp}^2-{\cTp}^2    \right)\pdv{^2 u_l}{x_i \partial{x_l}} + {\cTp}^2 \pdv{^2 u_i}{x_l \partial{x_l}} = 0, 
$$
where a displacement field oscillates sinusoidally with a frequency $\omega$ [i.e., $\fn{u_i}{\vect{x},t} = \fn{u_i}{\vect{x}}e^{-i\omega t}$], indices span integers between $1$ and $d$, and the Einstein summation is implied. 
Here, $\cLp$ and $\cTp$ represent the longitudinal and transverse wave speeds\footnote{Henceforth, `wave speeds' always refer to the phase speeds, because the term `phase' is reserved for a constituent material in this paper.} in phase $p$, respectively, and they are given as 
$$
 {\cLp}^2 \equiv [K_p +2(1-1/d)G_p]/\rho_p,~~~{\cTp}^2 \equiv G_p/\rho_p,  
$$
where $K_p$ and $G_p$ are the bulk modulus and the shear modulus of phase $p$, respectively. 
For a frequency $\omega$, the corresponding longitudinal and transverse wavenumbers for elastic waves in phase $p$ (=1,2) are denoted by
\begin{equation}
k_{_{{_L}_p}} \equiv \omega / \cLp, ~\text{and} ~ k_{_{T_p}} \equiv \omega / \cTp , \label{eq:wavenumbers}
\end{equation}
respectively.
Henceforth, we take ``reference'' and ``polarized'' phases to be phase 1 and 2, respectively; see Refs. \cite{Torquato1997, Torquato_RHM}.

Formulas for the effective dielectric constnat $\fn{\epsilon_e}{k_1}$ and the effective elastic moduli $\fn{K_e}{\kL}$ and $\fn{G_e}{\kL}$ also lead to estimates of the effective wave characteristics, including effective wave speeds $c_e$ and attenuation coefficient $\gamma_e$.
For electromagnetic and elastic waves, the analogous quantities are given by 

\begin{equation}\label{eq:effective_wave_propagations}
\begin{array}{lr}
c_e /c_1 + i \gamma_e /c_1= \sqrt{\epsilon_1/\epsilon_e},\\
c_e^L + i \gamma_e^L = \sqrt{[K_e +2(1-1/d)G_e]/\rho_e},\\
c_e^T + i \gamma_e^T = \sqrt{G_e/\rho_e},
\end{array}
\end{equation}
where $c_1$ is the wave speed of electromagnetic waves in the reference phase, and $\rho_e$ is the effective mass density ($\rho_e=\rho_1 = \rho_2$).
Note that, for the scaled attenuation coefficients $\gamma_e / c_e$, $\gamma_e^L / c_e^L$, and $\gamma_e^T / c_e^T$, a quantity $\exp[{-2\pi \gamma_e / c_e}]$  represents the factor by which the amplitude of the incident wave is attenuated for a period of time $2\pi/ \omega$.

\section*{Results}
	We first derive the microstructure-dependent formulas for the effective dynamic dielectric constant, bulk modulus, and shear modulus that apply from infinite wavelengths down to intermediate
wavelengths. 
	Then we use these estimates to establish cross-property relations between them by eliminating a common microstructural parameter among them.
	Using these formulas, we estimate the effective elastic wave characteristics and cross-property relations for four different 3D models of disordered two-phase dispersions, including two typical nonhyperuniform ones.
	Finally, we discuss how to employ the newly established cross-property relations in designing multifunctional materials. 

\subsection*{Microstructure-dependent approximation formulas}

\subsubsection*{Effective dielectric constant}\label{sec:effective_dielectric}
We begin with the strong-contrast approximation formula for $\fn{\epsilon_e}{k_1}$ in the long-wavelength regime ($k_1 \ell \ll 1$)  derived by Rechtsman and Torquato \cite{Rechtsman2008}.
Here, we modify this approximation so that it becomes valid down to intermediate wavelengths ($k_1 \ell \lesssim 1$). 

For macroscopically isotropic media, this formula depends on a functional $\fn{A_2}{Q}$ involving $\spD{\vect{Q}}$ \cite{Rechtsman2008}:
	\begin{equation}\label{eq:strong-contrast_dielectric_2pt}
	\frac{\fn{\epsilon_e}{k_1}}{\epsilon_1} = 1 - \frac{d \beta {\phi_2}^2}{\fn{A_2}{k_1} \beta + \phi_2 (\beta\phi_2 -1)},  
	\end{equation}
	where $k_1$ is the wavenumber of the electromagnetic waves in the reference phase (phase 1), $d$ is the dimension, $\beta \equiv \qty(\epsilon_2 - \epsilon_1)/[\epsilon_2 +(d-1)\epsilon_1]$ is the dielectric ``polarizability,'' and $\epsilon_1$ and $\epsilon_2$ are the dielectric constants of phases 1 and 2, respectively.
Here we modify the functional $\fn{A_2}{Q}$ stated in Ref. \cite{Rechtsman2008} by including contribution from spatial distribution of incident electric fields, e.g., $\exp(ik_1 x)$:  
\begin{equation}
\fn{A_2}{Q} = -\frac{(d-1)\pi}{2^{d/2}\fn{\Gamma}{d/2}} \fn{F}{Q} \label{eq:A2},
\end{equation}
where $\fn{F}{Q}$ is what we call the {\it attenuation function}; see Eq. \ref{eq:F_function_direct}.
Physically, the attenuation function $\fn{F}{Q}$ incorporates the contributions from all diffracted waves due to single, elastic scattering events when the wavenumber of the incident waves is $Q$.
The reader is referred to {\it\color{blue}Materials and Methods} and Sec. IV in the {\it\color{blue}SI} for a derivation of $\fn{F}{Q}$.
Numerical simulations of $\fn{\epsilon_e}{k_1}$ reported in Sec. V of the {\it\color{blue}SI} validate the high-predictive power of Eqs. \ref{eq:strong-contrast_dielectric_2pt} and \ref{eq:A2} for a wide range of incident wavelengths, which popular approximation schemes \cite{sihvola_electromagnetic_1999, markel_maxwell_2016} cannot predict, as shown in the {\it\color{blue}SI}.

For statistically isotropic media in three dimensions, the attenuation function can be rewritten as
\begin{align}
\Im[\fn{F}{Q}] =& - \frac{Q}{2(2\pi)^{3/2}} \int_0^{2Q} q\spD{q}\dd{q}, \label{eq:ImF_3D} \\
\Re[\fn{F}{Q}] = & -\frac{2 Q^2}{\pi } 
\mathrm{p.v.}  \int_{0}^{\infty} \dd{q}\frac{1}{q(Q^2-q^2)}\Im[\fn{F}{q}],\label{eq:ReF_3D}
\end{align}
where $\mathrm{p.v.}$ stands for the Cauchy principal value.
We elaborate on how to compute the attenuation function in Sec. IV of the {\it\color{blue} SI}.

\subsubsection*{Effective elastic moduli}\label{sec:effective_elastic}
We extract the approximations for $\fn{K_e}{\kL}$ and $\fn{G_e}{\kL}$ in the long-wavelength regime $(\kL \ell \ll 1)$ from the strong-contrast expansions (Eqs. \ref{eq:expansion_bulk_modulus} and \ref{eq:expansion_shear_modulus}) at the two-point level:  
\begin{align}
\frac{\fn{K_e}{\kL}}{K_1} &= 1 -\frac{\kappa  {\phi _2}^2 [1+ 2(1-1/d) G_1/K_1)]}{ \qty[\fn{C_2}{\kL}+\phi _2 \left(\kappa  \phi _2-1\right)]}, \label{eq:two-point_Ke}\\
\frac{\fn{G_e}{\kL}}{G_1} &= 1-\frac{(d+2)\mu  {\phi _2}^2 [1 + 2(1-1/d)G_1 / K_1]}{2 (1+2 G_1/K_1) \qty[\fn{D_2}{\kL}+\phi_2 (\mu  \phi _2-1) ]}
\label{eq:two-point_Ge},
\end{align}
where $\fn{C_2}{\kL}$ and $\fn{D_2}{\kL}$ are microstructural parameters that depend on $\kL$ because $\kT = (\cL/\cT ) \kL$. 
These strong-contrast approximations are also valid in the long-wavelength regime ($\kL \ell \ll 1 $).
However, we modify them so that they are valid in the intermediate-wavelength regime ($\kL \ell \lesssim 1$) by using the following modified forms of the microstructure-dependent parameters: 
\begin{align}\label{eq:C2}
\fn{C_2}{\kL}& = \frac{\pi}{2^{d/2} \fn{\Gamma}{d/2}} \fn{F}{\kL}\kappa, \\
\fn{D_2}{\kL}& = \frac{\pi}{2^{d/2} \fn{\Gamma}{d/2}} \frac{d\cL^2 \fn{F}{\kT} + 2\cT^2 \fn{F}{\kL}}{d\cL^2 + 2\cT^2}\mu, \label{eq:D2}
\end{align}
where $\fn{F}{Q}$ is the attenuation function, defined in Eq. \ref{eq:F_function_direct}.
The reader is referred to {\it\color{blue}Materials and Methods} for a derivation of these relations. Computer simulations reported in Sec. V of the {\it\color{blue}SI} verify that these modified formulas accurately predict microstructure-dependence of $\fn{K_e}{\kL}$ and $\fn{G_e}{\kL}$ down to intermediate wavelengths, where conventional approximation schemes \cite{kerr_scattering_1992} are no longer valid.

Note that the approximations (Eqs. \ref{eq:two-point_Ke} and \ref{eq:two-point_Ge}) are conveniently written in terms of the wavenumber $\kL$ associated with the reference phase. 
Furthermore, this wavenumber is directly proportional to the frequency $\omega$ (Eq. \ref{eq:wavenumbers}) and more suitable to describe microstructural information rather than the temporal quantity $\omega$. 
	For these reasons, we henceforth use the longitudinal wavenumber $\kL$, instead of $\omega$ or $\kT$, as an independent variable for the effective elastic properties.

\subsubsection*{Static limit} \label{sec:static_limit}
In the long-wavelength limit ($Q \to 0$), the attenuation function vanishes as $\fn{F}{Q} \sim \order{Q^2}$, which enables us to recover the previous corresponding static results \cite{Torquato1985, Torquato1997} in which the effective properties are real-valued and identical to the Hashin-Shtrikman bounds on $\epsilon_e$, $K_e$, and $G_e$ \cite{Hashin1962, Torquato_RHM, Milton_TheoComposites}, respectively.
Note that these static limits are also identical to the static Maxwell-Garnet approximations; see Sec. V in the {\it\color{blue}SI}.

\subsubsection*{Long-wavelength regime}
The effective dynamic properties $\fn{\epsilon_e}{k_1}$, $\fn{K_e}{\kL}$, and $\fn{G_e}{\kL}$ are generally complex-valued, implying that the associated waves propagating through this medium are attenuated.
Such attenuation occurs due to scattering from the inhomogeneity even when both phases are dissipationless (i.e., real-valued, as we assume throughout this study).

In the long-wavelength regime ($Q\ell \ll 1$), we now demonstrate that attenuation is stronger in a nonhyperuniform medium than in a hyperuniform one by comparing their leading-order terms of the effective attenuation coefficients. 
The imaginary parts of effective properties and, importantly, the associated effective attenuation coefficients $\gamma_e$ are approximately proportional to $\Im[\fn{F}{Q}]$, which is easily evaluated from the spectral density by using Eq. \ref{eq:ImF_3D}.
When the spectral density follows the power-law form $\tilde{\chi}_{_V}(Q) \approx a_1 + a_2 Q^\alpha$, where for nonhyperuniform media $a_1>0$ and hyperuniform media $a_1=0,~a_2>0$, and $\alpha>0$,  $\gamma_e$ in the limit $Q\to 0$ is given by  
\begin{equation*}
\gamma_e \sim
\begin{cases}
Q^3, &\text{nonhyperuniform},\\
Q^{3+\alpha}, &\text{hyperuniform}.
\end{cases}
\end{equation*}
Thus, hyperuniform media are 
less dissipative than nonhyperuniform
media due to the complete suppression (in the former) of single scattering events in the long-wavelength limit. 
Note for stealthy hyperuniform media [$\spD{Q}=0$ for $0<Q<Q_\text{upper}$], which is the strongest form of hyperuniformity, $\gamma_e=0$ in the same limit; see the section {\bf Transparency conditions} for details.

\subsubsection*{Transparency conditions}\label{sec:trans}

Our formulas [Eqs. \ref{eq:two-point_Ke} and \ref{eq:two-point_Ge}] predict that heterogeneous media can be transparent for elastic waves [$\fn{\gamma_e^{L,T}}{\kL}=0$] if
\begin{equation}\label{eq:tranparent regime}
\Im[\fn{F}{\kL}]=\Im[\fn{F}{\kT}]=0.
\end{equation}
Physically, these conditions imply that single scattering events of elastic waves at the corresponding frequency are completely suppressed.
For stealthy hyperuniform media that satisfy $\spD{Q}=0$ in $0< Q < Q_\mathrm{upper}$, the transparency conditions \ref{eq:tranparent regime} are simply given as 
\begin{equation*}
0 <\kL a< (\cT/\cL)Q_\mathrm{upper} a/2,
\end{equation*} 
where $\cT/\cL = \sqrt{(1-2\nu_1)/[2(1-\nu_1)]}$, where $\nu_1$ is the Poisson ratio of phase 1. 
For electromagnetic waves, the condition \ref{eq:tranparent regime} is simplified as $\Im[\fn{F}{k_1}]=0$.
Thus, the aforementioned stealthy hyperuniform media are transparent to the electromagnetic waves ($\gamma_e=0$) for $0<k_1 a < Q_\mathrm{upper} a/ 2$; see Sec. V in the {\it\color{blue}SI}. 
We will make use of these interesting properties in the sections {\bf Effective elastic wave characteristics} and {\bf Conclusions and Discussions.}

\subsection*{Models of dispersions}\label{sec:models}

\begin{figure}[h!]
\includegraphics[width =0.95\linewidth]{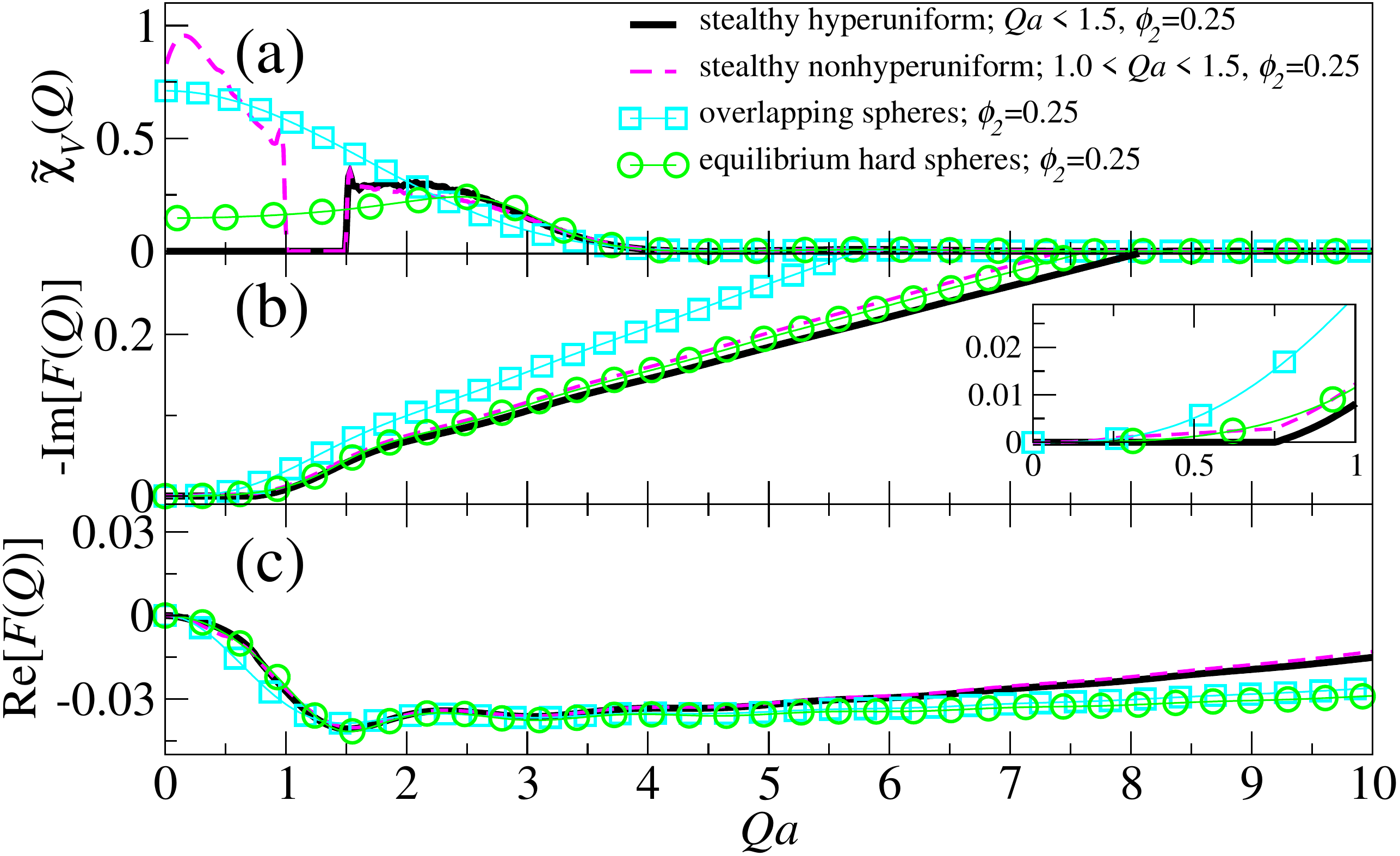}
\caption{Evaluation of $\spD{Q}$ (a) and the attenuation function $\fn{F}{Q}$ (b,c) for four different models of 3D dispersions: stealthy hyperuniform dispersions, stealthy nonhyperuniform dispersions, overlapping spheres, and equilibrium hard spheres.
The inset in (b) is a magnification of the larger panel. 
Since all of these dispersions are composed of spheres of radius $a$, their microstructures resemble one another at the small length scales $r<a$, indeed all of the spectral densities effectively collapse onto a single curve for $Qa \gg 5$. 
\label{fig:F functions}
}
\end{figure}

\begin{figure*}[th]
\includegraphics[width=0.8\textwidth]{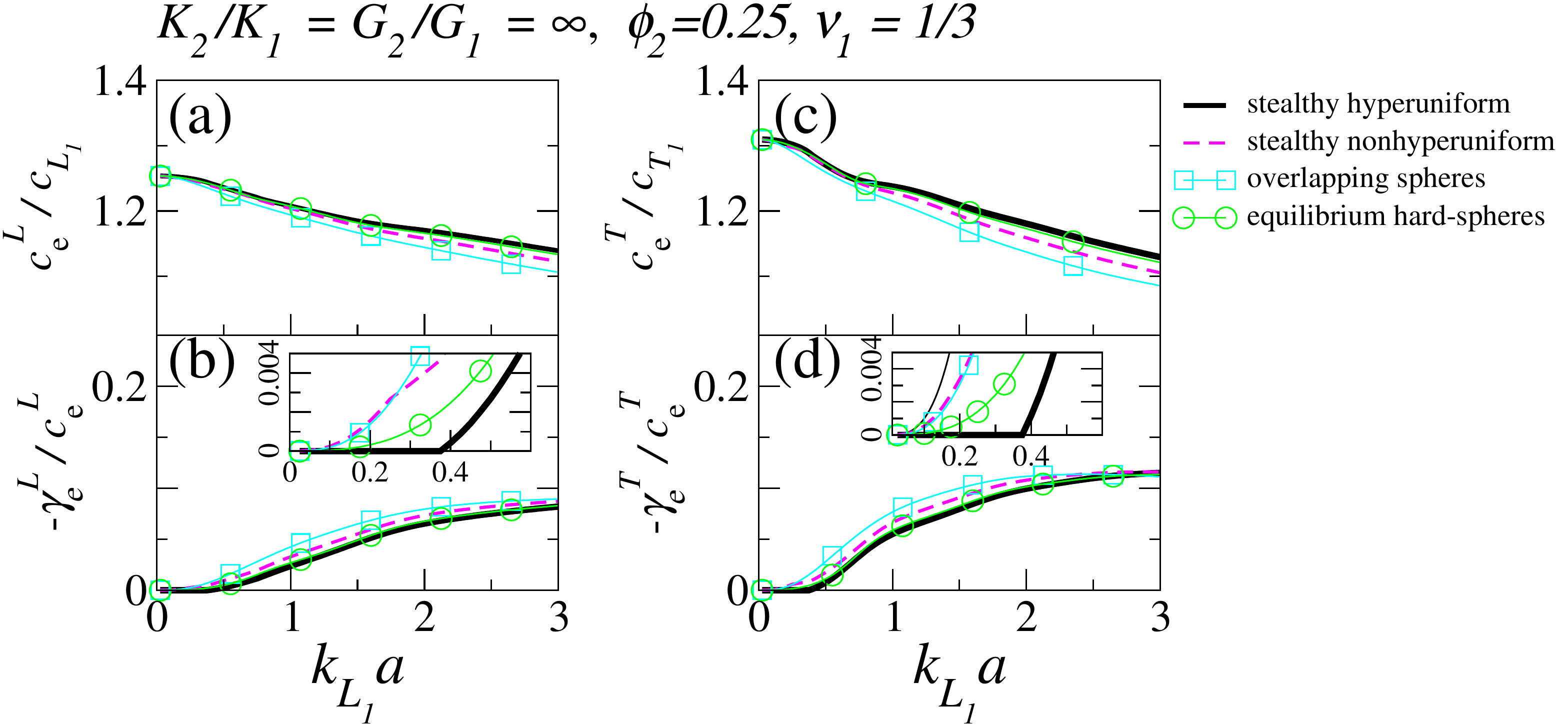}
\caption{Estimates of scaled effective elastic wave characteristics for 3D dispersions of rigid spheres of radius $a$ in a compressible matrix phase with Poisson ratio $\nu_1 = 1/3$ (i.e., $K_2/K_1 = G_2/G_1 = \infty$).
Here, $\kL$ is the wavenumber of longitudinal waves in reference phase (phase 1), and $\cL$ and $\cT$ are elastic wave speeds of longitudinal and transverse waves, respectively, in phase 1. 
(a,c) Effective wave speeds and (b,d) effective attenuation coefficients are plotted in terms of $\kL a$.
The insets in (b) and (d) are magnifications of the larger panels, respectively.
In these insets, we see that stealthy hyperuniform dispersions have a zero-$\gamma_e^{L,S}$ regime in $\kL a < 0.375$, but others do not, which vividly demonstrates that $\gamma_e^{L,S}$ can be engineered by the spatial correlations of composites.  
\label{fig:IncompressibleDispersions}}
\end{figure*}

    We investigate four different 3D models of disordered dispersions of identical spheres of radius $a$ with $\phi_2=0.25$.
    These models include two typical disordered ones (overlapping spheres and equilibrium hard spheres) and two exotic disordered ones (stealthy hyperuniform dispersions and stealthy nonhyperuniform dispersions).

	{\it Overlapping spheres} are systems composed of spheres whose centers are spatially uncorrelated \cite{Torquato_RHM}. 
	At $\phi_2=0.25$, this model will not form macroscopically large clusters, since it is well below the percolation threshold $\phi_2\approx 0.29$ \cite{Rintoul1997}.
	To compute its attenuation function $\fn{F}{Q}$, we use the closed-form expression of its $\fn{\chi_{_V}}{r}$, given in Ref. \cite{Torquato_RHM}; see also Sec. II in the {\it\color{blue}SI}.

	{\it Equilibrium hard spheres} are systems of nonoverlapping spheres in the canonical ensemble \cite{Hansen_statmechReference}.
	To evaluate its $\fn{F}{Q}$, we use the spectral density $\spD{Q}$ that is obtained from Eq. \ref{eq:chik_mono_packing} and the Percus-Yevick approximation; see Ref. \cite{Torquato1985b} and Sec. II in the {\it\color{blue} SI.}

	{\it Stealthy hyperuniform dispersions} are defined by $\spD{\vect{Q}}=0$ for $0<\abs{\vect{Q}}\leq Q_\mathrm{upper}$.	
    We numerically generate them via the collective-coordinate optimization technique \cite{Uche2004, Batten2008, Zhang2015}; see {\it\color{blue} Methods} for details.
    Each of these obtained systems consists of $N=10^3$ spheres and satisfies $\spD{Q} =0 $ for $Qa < 1.5$.
    In order to compute $\fn{F}{Q}$, we obtain $\spD{Q}$ numerically from Eq. \ref{eq:chik_mono_packing}; see {\it\color{blue}Methods} and Sec. III in the {\it\color{blue}SI}.

	{\it Stealthy nonhyperuniform dispersions} are defined by $\spD{Q}=0$ for $Q_\mathrm{lower} < Q< Q_\mathrm{upper}$.
	We numerically find realizations of these systems ($N=10^3$ and $\spD{Q} =0 $ for $1.0< Qa < 1.5$) via the collective-coordinate optimization technique \cite{Uche2004, Batten2008, Zhang2015}; see {\it\color{blue} Methods}.	
	Its spectral density is obtained in the same manner as we did for the stealthy hyperuniform dispersions.

	Values of the complex-valued attenuation function $\fn{F}{Q}$ (see Eqs. \ref{eq:ImF_3D} and \ref{eq:ReF_3D}) for the four aforementioned models of dispersions are presented in Fig. \ref{fig:F functions}. 
	Their imaginary parts are directly obtained from the spectral density based on Eq. \ref{eq:ImF_3D}.
	The associated real parts are then computed from an approximation of Eq. \ref{eq:ReF_3D}; see Sec. IV in the {\it\color{blue} SI}. 
	For various types of dispersions, while the values of $\spD{Q}$ up to intermediate wavenumbers ($Qa \lesssim 3$) are considerably different from one another, all of the curves approximately collapse onto a single curve for $Qa \gg 5$; see Fig. \ref{fig:F functions}(a).

\subsection*{Effective elastic wave characteristics} \label{sec:numerical results}

We now investigate the aforementioned effective elastic wave characteristics of four different models of 3D dispersions, using the strong-contrast approximations (Eqs. \ref{eq:two-point_Ke} and \ref{eq:two-point_Ge}).
In striking contrast to the other models, stealthy hyperuniform dispersions are transparent for both longitudinal and transverse elastic waves down to a finite wavelength.
This result clearly demonstrates that it is possible to design disordered composites that exhibit nontrivial attenuation behaviors by appropriately manipulating their spatial correlations.  

We first determine phase elastic moduli of the aforementioned four models of composites. 
Since this parameter space of phase moduli is infinite, we consider two extreme cases: a compressible matrix phase (phase 1) with a Poisson ratio $\nu_1=1/3$ that contains a rigid dispersed phase (phase 2), i.e., $K_2/K_1 = G_2/G_1 = \infty$ (Fig. \ref{fig:IncompressibleDispersions}), and a compressible matrix with $\nu_1=-0.1$ that contains cavities, i.e., $K_2/K_1 = G_2/G_1 = 0$ (Sec. VI in the {\it\color{blue}SI}).
Investigating these two extreme cases will still provide useful insight into the wave characteristics in intermediate regimes of phase moduli.
	While the Poisson ratio of the compressible matrix phase can take any value in the allowable interval of $-1\leq \nu_1 <1/2$, we examine two different values of $\nu_1=1/3$ (typical of many materials), and $-0.1$.
Negative Poisson ratio (``auxetic'') materials laterally dilate (contract) in response to axial elongation (contraction) \cite{Milton1992}. 
    While we present the estimated effective elastic moduli up to $\kL a =3$, our approximations are, in principle, valid down to the intermediate-wavelength regime ($\kL a <1.5$).

\begin{figure*}[t]
\begin{center}
\includegraphics[width=0.9\textwidth]{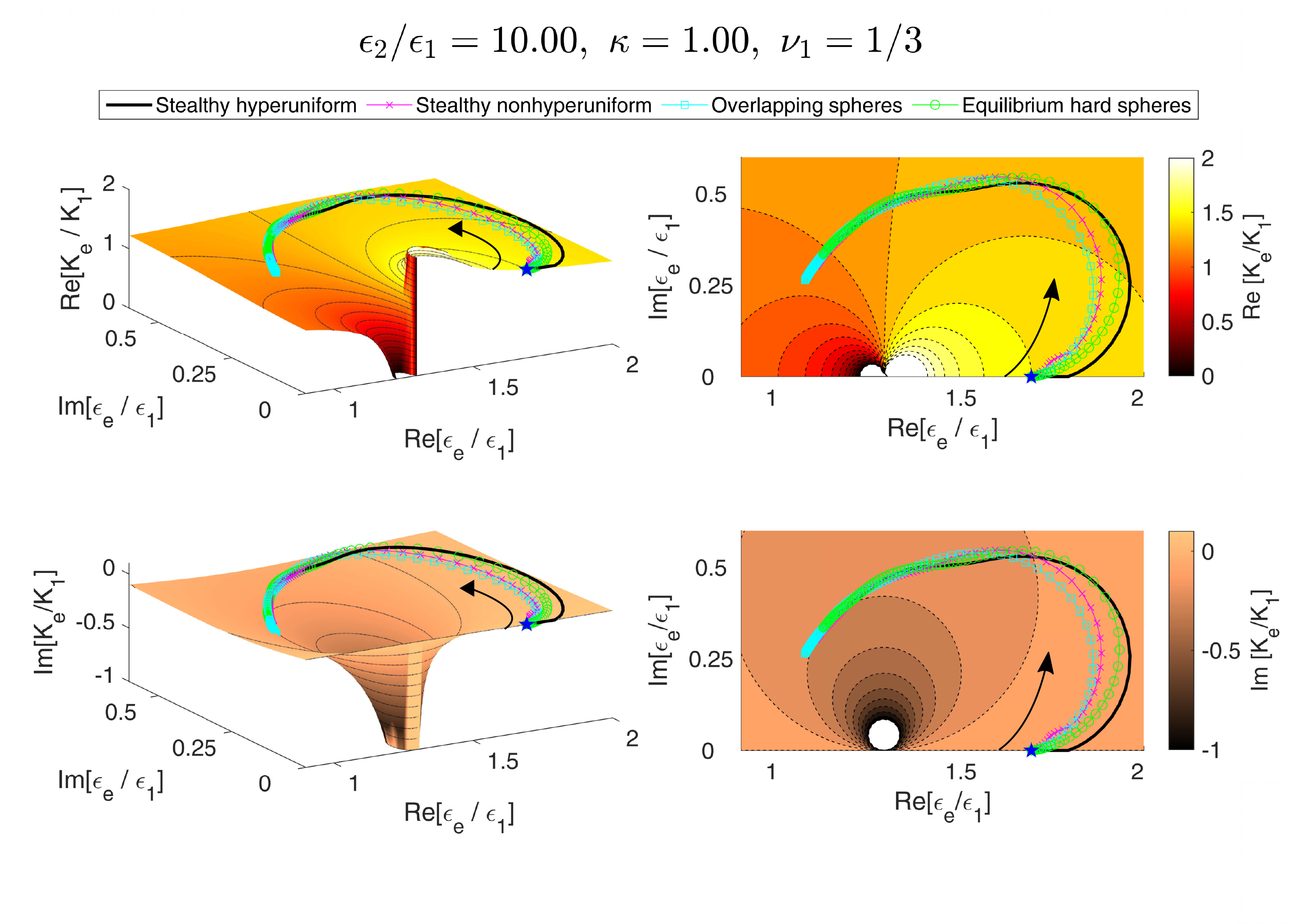}\vspace{-20pt} 
\end{center}
\caption{Cross-property relation between the effective dielectric constant $\epsilon_e$ and the effective bulk modulus $K_e$ for the four models of 3D dispersions with $\phi_2=0.25$, each of which consists of a compressible matrix with $\nu=1/3$ and incompressible inclusions (see Fig. \ref{fig:IncompressibleDispersions}) and the ratio of phase dielectric constants is $\epsilon_2/\epsilon_1=10$.
Left: surface plots, evaluated from  Eq. \ref{eq:Ee2Ke}, represent the surface on which $\epsilon_e$ and the real (the upper panels) and imaginary (the lower panels) parts of $K_e$ of any composites with the prescribe phase properties must lie.
Contour lines (black dotted lines) are at level spacing 0.1.
Right: contour plots are the top views of the surface plots on the left panels. 
We note that the surface plots for other values of the Poisson ratio $\nu_1$ and the contrast ratio $\epsilon_2/\epsilon_1$ are qualitatively similar except for the position of a simple pole; see Figs. S6-S7 in the {\it\color{blue}SI}. 
On these surfaces, the locus of points (solid lines with/without markers) represents effective properties of each dispersion model as the wavenumber $k_1$ increases from $k_1 a=0$ to $k_1 a = 5$ (in the directions of arrows).
Note that all the models attain the Hashin-Shtrikman bounds (blue stars) at $k_1 a=0.$  
\label{fig:cross-property}
}
\end{figure*}

	We estimate the scaled effective wave propagation properties of the models of 3D dispersions considered in Fig. \ref{fig:F functions}.
	For each of the aforementioned cases of phase properties, four different models have similar effective wave speeds but significantly different attenuation coefficients.
	Instead $\fn{c_e^L}{\kL}$ and $\fn{c_e^T}{\kL}$ depend largely on the phase properties.
For rigid dispersed phase (Fig. \ref{fig:IncompressibleDispersions}), the effective wave speeds $c_e^{L,T}$ are generally faster than those in phase 1 but tend to decreases with $\kL$ at most frequencies.
By contrast, when the dispersed phase consists of cavities, the wave speeds are slower than those in phase 1 and increases with $\kL a$ from $\kL a \approx 1$; see Sec. VI in the {\it\color{blue} SI}.

In both cases shown in Figs. \ref{fig:IncompressibleDispersions} and S5, stealthy hyperuniform dispersions are transparent to both longitudinal and transverse waves in $0<\kL a \lesssim 0.4$, as predicted in Eq. \ref{eq:tranparent regime}. 
Such composites can be employed to design of low-pass filters for elastic as well as electromagnetic waves.
By contrast, the stealthy nonhyperuniform dispersions do not attain zero attenuation at any finite wavelength because these systems can suppress scatterings at only specific directions.

\subsection*{Cross-property relations}\label{sec:multifunctional}
    It is desired to design composites with prescribed elastic and electromagnetic wave characteristics, as schematically illustrated in Fig. \ref{fig:schem}.
    The rational design of such multifunctional characteristics can be greatly facilitated via the use of cross-property relations for these different effective properties, which we derive here.

\begin{table*}
\caption{\label{tab:Ge}
Evaluation of the effective shear moduli $G_e/G_1$ for dispersions of spheres of radius $a$ from the strong-contrast approximation (Eq. \ref{eq:two-point_Ge}) and the cross-property relation (Eq. \ref{eq:Ee2Ge}).
\label{eq:cross-property_epsilon2G}}
\begin{tabular}{c|c| c l | c l | c l | c l}
\hline
\multirow{2}{*}{3D models} & \multirow{2}{*}{$\kL a$} &
\multicolumn{2}{c|}{\multirow{2}{*}{$\fn{\epsilon_e}{\kL}/\epsilon_1$}} & \multicolumn{2}{c|}{\multirow{2}{*}{$\fn{\epsilon_e}{\cL \kL / \cT}/\epsilon_1$}} & 
\multicolumn{4}{c}{$\fn{G_e}{\kL} / G_1$} \\
\cline{7-10}
& & \multicolumn{2}{c|}{} & \multicolumn{2}{c|}{} & \multicolumn{2}{c|}{From Eq. \ref{eq:Ee2Ge}} & \multicolumn{2}{c}{From Eq.  \ref{eq:two-point_Ge} }\\
\hline
\multirow{2}{*}{\parbox{0.2\textwidth}{Overlapping\\spheres}} & 0.3 &1.724& $+i$8.239$\times 10^{-3}$&1.808& $+i$7.455$\times 10^{-2}$&1.650& $-i$ 2.941$\times 10^{-2}$&1.650& $-i$ 2.941$\times 10^{-2}$\\
    & 0.5 & 1.776& $+i$4.135$\times10^{-2}$&1.843& $+i$3.420$\times 10^{-1}$&1.585&$-i$ 8.904$\times 10^{-2}$&1.585& $-i$ 8.904$\times 10^{-2}$ \\
    \hline
\multirow{2}{*}{\parbox{0.2\textwidth}{Equilibrium\\hard spheres}} & 0.3 & 1.708& $+i$1.723 $\times 10^{-3}$ &1.758& $+i$ 1.673 $\times 10^{-3}$ &1.676& $-i$ 8.132 $\times 10^{-3}$&1.676& $-i$ 8.139 $\times 10^{-3}$ \\
    & 0.5 &   1.737& $+i$ 8.938 $\times 10^{-3}$&1.888& $+i$1.240 $\times 10^{-1}$&1.618&$-i$ 3.614 $\times 10^{-2}$&1.618&$-i$ 3.617 $\times 10^{-2}$ \\
    \hline
\multirow{2}{*}{\parbox{0.2\textwidth}{Stealthy\\hyperuniform dispersions }} & 0.3 & 1.704& $+i$1.510$\times 10^{-27}$&1.745& $+i$3.477$\times 10^{-27}$&1.683& $-i$ 1.894$\times 10^{-27}$&1.683& $-i$ 1.894$\times 10^{-27}$\\
    & 0.5 & 1.727& $+i$2.739$\times 10^{-27}$ &1.913& $+i$9.130$\times 10^{-2}$&1.615&$-i$ 2.469$\times 10^{-2}$&1.615& $-i$ 2.469$\times 10^{-2}$ \\
    \hline
\multirow{2}{*}{\parbox{0.2\textwidth}{Stealthy\\nonhyperuniform dispersions  }} & 0.3 &1.716& $+i$9.594$\times 10^{-3}$&1.755& $+i$4.643$\times 10^{-2}$&1.675& $-i$ 2.302$\times 10^{-2}$&1.675& $-i$ 2.302$\times 10^{-2}$ \\
    & 0.5 & 1.742& $+i$3.726$\times 10^{-2}$ &1.875& $+i$1.908$\times 10^{-1}$&1.611&$-i$ 5.638$\times 10^{-2}$&1.611& $-i$ 5.638$\times 10^{-2}$ \\
    \hline            
\end{tabular}
	\addtabletext{
    Phase moduli are identical to those considered in Fig. \ref{fig:IncompressibleDispersions} (i.e., $K_2/K_1=G_2/G_1=\infty$ and $\nu_1 =1/3$), and $\epsilon_2/\epsilon_1 = 10$.
    Here, $\cL$ and $\cT$ are the longitudinal and transverse elastic wave speeds in the reference phase (phase1), respectively, and $\kL$ is the wavenumber of longitudinal elastic waves in phase 1. 
Those two formulas give consistent values for $G_e$ at two distinct wavenumbers which vividly demonstrates that $G_e$ can be indirectly evaluated from the wavenumber-dependent $\fn{\epsilon_e}{k_1}$. }
\end{table*}

We first obtain a cross-property relation between the effective dynamic bulk modulus and effective dynamic dielectric constant from Eqs. \ref{eq:strong-contrast_dielectric_2pt} and \ref{eq:two-point_Ke} by eliminating $\fn{F}{Q}$ between them:    
\begin{align}\label{eq:Ee2Ke}
\frac{\fn{K_e}{\kL}}{K_1} =& 1+\frac{6 \beta  \kappa \left(\nu_1-1\right) \phi _2 }{\nu _1+1} \\
&\times  
\frac{\fn{\epsilon_e}{\kL}/\epsilon_1 -1}{2 \beta +\kappa+\left(3 \beta  \kappa  \phi _2-2 \beta -\kappa \right) \fn{\epsilon _e}{\kL} /\epsilon_1 } \nonumber ,
\end{align}
where the effective properties $K_e$ and $\epsilon_e$ must be at the same wavenumber (i.e., $\kL = k_1$) but possibly at different frequencies, as illustrated in Fig. \ref{fig:schem}.
	Remarkably, this cross-property relation depends only on the phase properties, regardless of the microstructures of composites. 
	Intuitively speaking, such cross-property relations can be established because the effective properties depend on the interference pattern of the associated waves, which are commonly determined by wavelengths and microstructures. 

	The real and imaginary parts of this cross-property relation (Eq. \ref{eq:Ee2Ke}) are separately represented in Fig. \ref{fig:cross-property} for the four models of 3D dispersions considered in Fig. \ref{fig:IncompressibleDispersions}.
	The surface plots on the left panels in Fig. \ref{fig:cross-property} depict the hypersurface on which any possible pairs of $\fn{\epsilon_e}{k_1}$ and $\fn{K_e}{\kL = k_1}$ of a composite must lie when its phase properties and $\phi_2$ are prescribed.
	The black dotted lines in the upper and lower panels are contour lines of $\Re[K_e/K_1]$ and $\Im[K_e/K_1]$ at level spacing 0.1, respectively.
	The right panels in Fig. \ref{fig:cross-property} represent the top views of the associated surface plots on the left panels.
	We note that the resulting surface plots have a simple pole at $\epsilon_e = \epsilon_\mathrm{pole}$ whose position is determined by phase properties and packing fraction $\phi_2$; see also Figs. S6-S7 in the {\it\color{blue}SI}.
	In Fig. \ref{fig:cross-property}, the locus of points (shown in solid lines) depicts the effective dielectric constants and bulk moduli of the four different models of 3D dispersions as a dimensionless wavenumber $k_1 a(=\kL a)$ varies from 0 to 5 along with the arrows.
	Since these points should lie on the surfaces as depicted in Fig. \ref{fig:cross-property}, one can indirectly determine the wavenumber-dependent $K_e$ by measuring $\epsilon_e$ at different wavenumbers (or frequencies), and vice versa.

Similarly, we can obtain cross-property relations that links $\epsilon_e$ to $G_e$ or $G_e$ to $K_e$. 
The former case is explicitly given as 
\begin{align}
&\frac{\fn{G_e}{\kL}}{G_1} = 1 - 
    15 (\nu_1 -1){\phi_2} \mu\times \label{eq:Ee2Ge} \\
   & \Bigg( (5\nu_1 -4)\Bigg\{ 
        \qty[ \phi_2 (1+2\mu) - \frac{1+2\beta}{\beta} ] + \frac{3\phi_2}{3\cL^2 + 2\cT^2} \times  \nonumber \\
   &~~  \qty[     \frac{3\cL^2} {\fn{\epsilon_e}{\cL \kL / \cT }/\epsilon_1 -1}
        + \frac{2\cT^2}{\fn{\epsilon_e}{\kL }/\epsilon_1 -1}]    
     \Bigg\}\Bigg)^{-1}, \nonumber 
\end{align}
which depends on values of the effective dielectric constant $\epsilon_e$ at two different wavenumbers $k_1 = \kL$ and $\cL\kL/\cT$, making it difficult to graphically depict this cross-property relation. 
Instead, we list in Table \ref{tab:Ge} values of $G_e$ that are computed from both Eqs. \ref{eq:two-point_Ge} and \ref{eq:Ee2Ge}. 
Furthermore, by combining Eqs. \ref{eq:Ee2Ke} and \ref{eq:Ee2Ge}, one can also establish cross-property relations that link the effective dielectric constant to the effective elastic wave characteristics, i.e., $c_e^{L,T}$ and $\gamma_e^{L,T}$.

\subsection*{Sound-absorbing and light-transparent materials}\label{sec:application}
To illustrate how our results can be applied for novel multifunctional material design, we engineer composites that are transparent to electromagnetic waves at infrared wavelengths (long wavelengths) but absorb sound at certain frequencies.
	Importantly, designing such materials is not possible by using standard approximations \cite{kerr_scattering_1992, sihvola_electromagnetic_1999, markel_maxwell_2016} and quasi-static cross-property relations \cite{Torquato1990, Gibiansky1993, Torquato2002, Torquato_RHM, Torquato2018_5, Milton1981b, Luis2007, Wang2016}. 
	Such engineered materials could be used as heat-sinks for central processing units (CPUs) and other electrical devices subject to vibrations or sound-absorbing housings \cite{lincoln_multifunctional_2019}.
	A similar procedure can be applied to design composites for exterior components of spacecraft \cite{Jiang2018} and building materials \cite{Guan2006}.   
	We will further discuss possible applications in {\bf Conclusions and Discussion}.  	

\begin{figure*}[h]
\includegraphics[width = 0.9\textwidth]{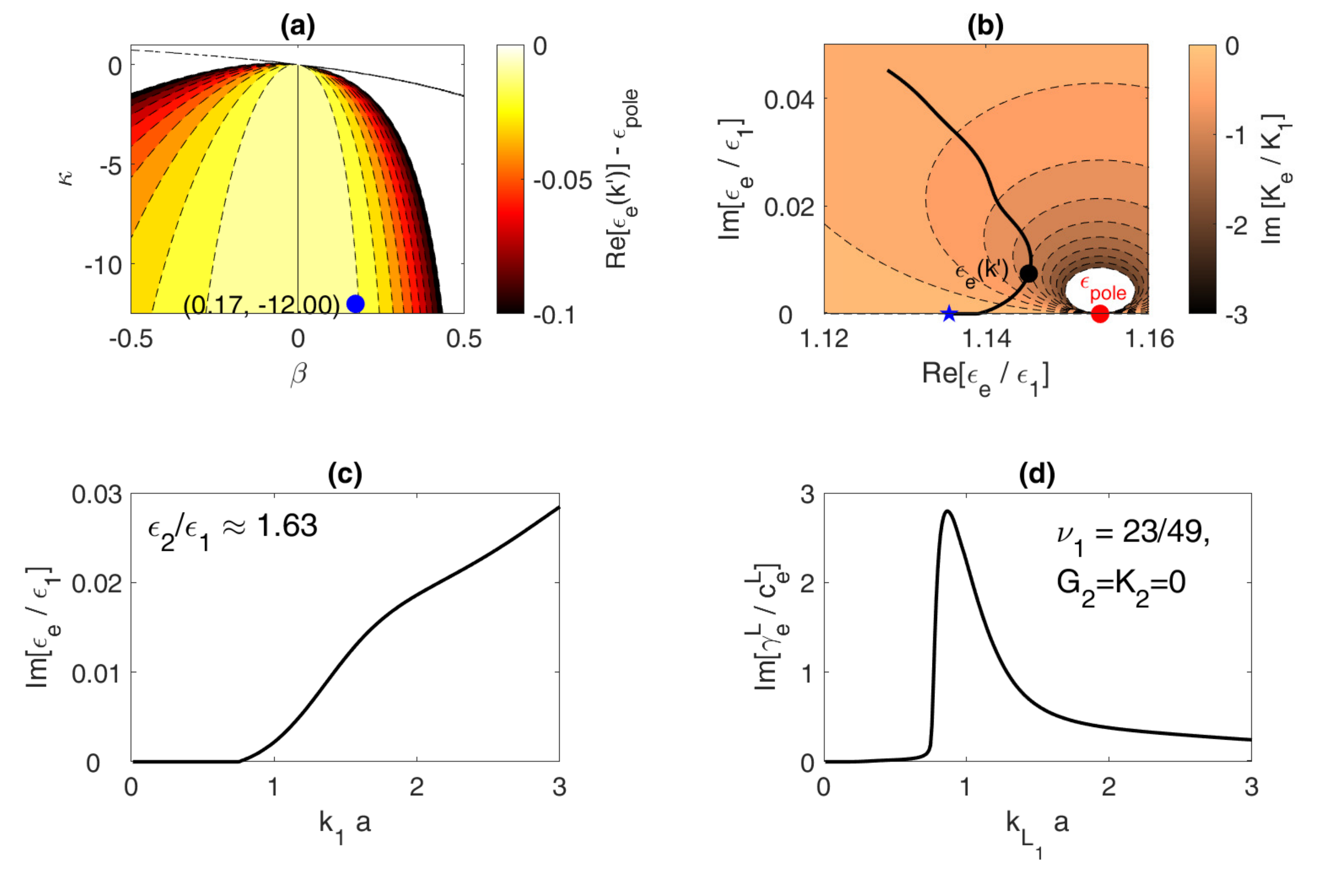}
\caption{Multifunctional design of materials that are transparent at infrared wavelengths but absorb sound at certain acoustic frequencies.
In order to attain such materials, we exploit exotic 3D stealthy hyperuniform dispersions.  
(a) Contour plot of the distance between the real part of the effective dielectric constant at a target frequency (the associated wavenumber is $k' a = 1.3$) and the simple pole $\epsilon_\mathrm{pole}$ as a function of the dielectric $\beta$ and bulk modulus $\kappa$ polarizabilities.
We choose $\beta = 0.17$ and $\kappa=-12$ (shown in a blue disk).
(b) The cross-property relation between $\epsilon_e/\epsilon_1$ and $\Im[K_e/K_1]$ for the chosen parameters $\beta=0.17$ and $\kappa = -12$.
The effective static dielectric constant is shown as a blue star.
The effective dielectric constant at the target frequency and the pole are shown in the black and red circles, respectively.
The associated composite consists of a nearly incompressible matrix phase with $\nu_1 = 23/49$ that containing cavities ($K_2=G_2=0$), and its contrast ratio of the dielectric constants is $\epsilon_2/\epsilon_1 \approx 1.63$.
(c) The imaginary part of the resulting effective dynamic dielectric constant as a function of the wavenumber $k_1 a$.
(d) The resulting scaled attenuation coefficient for the longitudinal elastic waves as a function of the longitudinal wavenumber $\kL$.
\label{fig:application}
}
\end{figure*}	

	We take advantage of the fact that stealthy hyperuniform dispersions are transparent down to a finite wavelength ($k_1 \leq Q_\text{upper}/2$); see Fig. \ref{fig:application}.
	We then find polarizabilities $\beta$ and $\kappa$ that result in high attenuation coefficient $\gamma_e^{L}$ at $\kL a= k' a \equiv 1.3$.
	This is achieved when $\fn{\epsilon_e}{k'}$ is close to the simple pole $\epsilon_\mathrm{pole}$ of Eq. \ref{eq:Ee2Ke}; see Fig. \ref{fig:application}(a). 	
	Figure \ref{fig:application}(b) shows the cross-property relation \ref{eq:Ee2Ke} with the chosen polarizabilites, i.e., $\beta=0.17$ and $\kappa=-12$.
	Phase properties corresponding to these polarizabilities are degenerate, and we choose $\epsilon_2/\epsilon_1\approx 1.63$, $\nu_1 = 23/49$, and $K_2=G_2=0$. 
	We see from Fig. \ref{fig:application}(c)-(d) that the resulting materials are indeed transparent to electromagnetic waves at long wavelengths but exhibit resonance-like attenuation of sound at $\kL a \approx 1.0$.

\section*{Conclusions and Discussion}\label{sec:discussion}

    We have obtained accurate approximations for the effective dynamic dielectric constant $\epsilon_e$ and the effective dynamic elastic moduli $K_e$ and $G_e$ of two-phase composites that depend on the microstructure via the spectral density $\spD{Q}$, which is easily computed theoretically/computationally or accessible via scattering experiments; see Eqs. \ref{eq:strong-contrast_dielectric_2pt}, \ref{eq:two-point_Ke}, and \ref{eq:two-point_Ge}.
	These formulas are superior in predicting these effective dynamic properties compared to commonly used approximations, such as Maxwell-Garnett and quasicrystalline approximations \cite{kerr_scattering_1992, sihvola_electromagnetic_1999, markel_maxwell_2016}, as verified by computer simulations reported in Sec. V of the {\it\color{blue}SI}.
	Unlike these conventional approximations, our formulas are accurate for a wide range of incident wavelengths for a broad class of dispersions.

    Using the approximations \ref{eq:strong-contrast_dielectric_2pt}, \ref{eq:two-point_Ke}, and \ref{eq:two-point_Ge}, we have shown that hyperuniform composites can have desirable attenuation properties both for electromagnetic and elastic waves.
    We analytically showed that hyperuniform media are less dissipative than nonhyperuniform ones in the section {\bf Long-wavelength regime}. 
    Remarkably, stealthy hyperuniform media are dissipationless (i.e., $\gamma_e$ = 0) down to a finite wavelength, as shown in Fig. \ref{fig:IncompressibleDispersions} and Figs. S2 and S5 in the {\it\color{blue}SI}.
    Such composites can be employed to low-pass filters for elastic and electromagnetic waves.

    Using Eqs. \ref{eq:strong-contrast_dielectric_2pt}, \ref{eq:two-point_Ke}, and \ref{eq:two-point_Ge}, we also established cross-property relations \ref{eq:Ee2Ke} and \ref{eq:Ee2Ge} that link the effective dynamic dielectric constant $\fn{\epsilon_e}{k_1}$ to the effective dynamic bulk modulus $\fn{K_e}{\kL}$ and shear modulus $\fn{G_e}{\kL}$, respectively.  
    Thus, when it is difficult to directly measure $K_e$ or $G_e$, they can be indirectly evaluated from these cross-property relations by measuring the wavenumber-dependent dielectric constants \cite{Surzhikov2008, Fursa2013}, as demonstrated in Fig. \ref{fig:cross-property} and Table \ref{tab:Ge}, and vice versa.
    For example, one can use them to indirectly determine physical/chemical properties for construction materials \cite{Surzhikov2008, Fursa2013} and oil-exploration \cite{carcione_cross-property_2007}. 
   
    Our cross-property relations also have important practical implications for the rational design of multifunctional composites \cite{Milton1981b, Torquato_RHM, Torquato2002, Luis2007, Wang2016, Torquato2018_5, lincoln_multifunctional_2019} that have the desired dielectric properties for a particular range of electromagnetic wavelengths and elastic properties for a certain range of elastodynamic wavelengths.
    The validation of our formulas via computer simulations justifies their use for the design of novel multifunctional materials without having to perform full-blown simulations. 
	In particular, we described how to
engineer a sound-absorbing composite that is transparent to light via our cross-property relations, which again could not be done using previous approximation formulas \cite{kerr_scattering_1992, sihvola_electromagnetic_1999, markel_maxwell_2016, Torquato1990, Gibiansky1993, Torquato2002, Torquato_RHM, Torquato2018_5, Milton1981b, Luis2007, Wang2016}. 
	This is done by exploiting the exotic structural properties of stealthy hyperuniform dispersions; see Fig. \ref{fig:application}.  
	Such engineered materials could be used as heat-sinks for CPUs and other electrical devices subject to vibrations because they enable radiative cooling while suppressing prescribed mechanical vibrations. 
	Another application is a sound-absorbing housing for an engine or a motor, which can efficiently convert cyclic noise into electric energy \cite{lincoln_multifunctional_2019} and allow radiative cooling. 
	It is natural to extend to the aerospace industry where low-frequency engine noise is prevalent  \cite{lincoln_multifunctional_2019}.
	A similar procedure can be applied to design composites with high stiffness that absorb electromagnetic waves at certain wavenumbers for use as exterior components of spacecraft \cite{Jiang2018} and building materials \cite{Guan2006}.    
    
With the aid of our microstructure-dependent formulas [Eqs. \ref{eq:strong-contrast_dielectric_2pt}, \ref{eq:two-point_Ke}, and \ref{eq:two-point_Ge}], one can employ {\it inverse-design} approaches \cite{Torquato2009_inverse} to design composites.
	We recall that inverse-design approaches enable one to prescribe the effective properties of composites and then find the microstructures that achieve them. 
	For example, one would  first prescribe the material phases and then
compute the desired effective properties (say, attenuation coefficients $\gamma_e$ for a given bandwidth) via the microstructure-dependent formulas.  
	Then, one backs out the corresponding spectral density from the attenuation function, which would correspond to
a particular microstructure, if realizable. 
	The latter can be constructed by using previously established Fourier-space construction techniques \cite{Uche2004, Batten2008, Zhang2015, Chen2017}. 
	Finally, one can generate simulated microstructures via modern fabrication methods, such as 3D printing \cite{wong_review_2012} or 2D photolithographic technologies \cite{zhao_assembly_2018}.
	The same inverse techniques also can be employed to design multifunctional composites using the cross-property relations. 
	Remarkably, such inverse approaches were not possible with previously known approximations, such as Maxwell-Garnett and quasicrystalline formulas \cite{kerr_scattering_1992, sihvola_electromagnetic_1999, markel_maxwell_2016} because they are independent of microstructures.

It is instructive to briefly discuss how to measure the wavenumber-dependent effective properties in experiments.
Here, for brevity, we focus on the dielectric constants because the same reasoning applies to the elastodynamic case (Eqs. \ref{eq:two-point_Ke} and \ref{eq:two-point_Ge}). 
Clearly, the property $\fn{\epsilon_e}{k_1}$ is identical to the frequency-dependent one because a wavenumber $k_1$ in the reference phase can be converted to a frequency $\omega$ via the dispersion relation of the reference phase [i.e., $\omega = \fn{\omega}{k_1}$]. 
The frequency-dependent effective dielectric constant $\fn{\epsilon_e}{\omega}$ can be measured via conventional techniques, such as perturbation methods (measuring changes in a resonance frequency of a resonator due to a specimen) or transmission techniques (measuring the transmission/reflection by a specimen); see Ref. \cite{tereshchenko_overview_2011} and references therein. 
However, when using cross-property relations, it is crucial to covert the independent variable of the effective properties from frequency $\omega$ to the associated wavenumbers ($k_1$ and $\kL$) according to the dispersion relations of the reference phase. 
	    
	While we primarily focused on three-dimensional two-phase media, our microstructure-dependent formulas (Eqs. \ref{eq:strong-contrast_dielectric_2pt}, \ref{eq:two-point_Ke}, and \ref{eq:two-point_Ge}) are valid for $d\geq 2$. 
	Furthermore, the cross-property relations (Eqs. \ref{eq:Ee2Ke} and \ref{eq:Ee2Ge}) can be extended to any dimension $d$ with minor modifications.
	
    Based on a previous study on the static case \cite{Gibiansky1997b}, it is relatively straightforward to generalize our microstructure-dependent formulas to composites whose dispersed phase is a piezoelectric (i.e., mechanical stress can induce an electric voltage in the solid material). 
    Such extensions can be profitably utilized in the optimal design of materials for elastic wave energy-harvesting to power small electrical devices \cite{Yuan_2019_energy-harvesting}.

\matmethods{

\subsection*{Derivation of Eq. \ref{eq:strong-contrast_dielectric_2pt}}

We begin with the original expression of the microstructure-dependent parameter $\fn{A_2}{Q}$ in the long-wavelength regime, derived in Ref. \cite{Rechtsman2008}:
\begin{align}
\fn{A_2}{Q} = &(d-1){Q}^2 \int \frac{i}{4}\qty(\frac{Q}{2\pi r})^{d/2-1} \Hankel{d/2-1}{Q r}\fn{\chi_{_V}}{\vect{r}}\dd{\vect{r}} \nonumber  \\
\equiv & -\frac{(d-1)\pi}{2^{d/2}\fn{\Gamma}{d/2}} \fn{\mathcal{F}}{Q}, \label{eq:A2_original}
\end{align}
where $Q$ is a wavenumber, $\fn{\Gamma}{x}$ is the gamma function, and $\Hankel{\nu}{x}$ is the Hankel function of the first kind of order $\nu$.
Here, the function $i/4\qty(Q/2\pi r)^{d/2-1} \Hankel{d/2-1}{Qr}$ in the integrand is the Green's function of the Helmholtz equation characterized by a wavenumber $Q$ in $d$-dimensional Euclidean space.
Note that the imaginary part of $\fn{\mathcal{F}}{Q}$ can be simplified as  
\begin{equation}
\Im[\fn{\mathcal{F}}{Q}] = - {Q}^d \spD{Q}/(2\pi)^{d/2}.\label{eq:F_original}
\end{equation}
The reader is referred to Sec. IV in the {\it\color{blue}SI} for discussion about physical interpretation of this quantity.

In order to extend the range of applicable wavelengths, we modify the microstructure-dependent parameter $\fn{A_2}{Q}$ to account for the spatial variation of the (external) incident waves, as in the Born approximation \cite{Jackson_TEXTBOOK_3rd}; see Eq. \ref{eq:A2}. 
The attenuation function $\fn{F}{Q}$ in Eq. \ref{eq:A2} is defined as  
\begin{align}
\fn{F}{Q} & =  \frac{-i \fn{\Gamma}{d/2}}{2\pi^{d/2}}{Q}^{d/2+1} \int \frac{\Hankel{d/2-1}{Q r}}{r^{d/2-1}} e^{-iQ \uvect{k} \cdot \vect{r}}\fn{\chi_{_V}}{\vect{r}} \dd{\vect{r}} \label{eq:F_function_direct}\\
& = 
-\frac{ \fn{\Gamma}{d/2}}{2^{d/2}\pi^{d+1}}{Q}^2 \int  \frac{\spD{\vect{q}}}{\abs{\vect{q}+Q\uvect{k}}^2 - Q^2}
\dd{\vect{q}},\label{eq:F_function_Fourier}
\end{align}
where $\uvect{k}$ is the unit wavevector in the direction of the incident waves.
Equation \ref{eq:F_function_Fourier} is obtained by applying the Parseval theorem to Eq. \ref{eq:F_function_direct}.
Importantly, comparison of the modified attenuation function (Eq.  \ref{eq:F_function_direct}) to Eq. \ref{eq:A2_original} reveals that the former has an additional factor $\exp(-i Q \uvect{k} \cdot \vect{r})$ in its integrand which account for the spatial variation of the incident waves.
This change enables us to include the contributions from all scattered waves at wavevectors $\vect{q}+Q\uvect{k}$; see Eq. \ref{eq:F_function_Fourier}.
We note that the modified approximation \ref{eq:strong-contrast_dielectric_2pt} with the attenuation function \ref{eq:F_function_Fourier} shows excellent agreement with numerical simulations; see Sec. V in the {\it\color{blue}SI}.

\subsection*{Dynamic strong-contrast expansions for the effective elastic moduli}
Elsewhere we derived exact strong-contrast expansions for these moduli through all orders in the ``polarizabilities.'' 
These expansions are, in principle, valid in the long-wavelength regime. 
Here it suffices to present their general functional forms when the effective stiffness tensor is isotropic: 
\begin{align}
\frac{K_e -K_1}{K_e + 2(d-1)G_1 /d} & = \frac{{\phi_2}^2 \kappa}{\phi_2 - \sum_{n=2}^\infty C_n}, \label{eq:expansion_bulk_modulus}\\
\frac{G_e -G_1}{G_e + \frac{[d K_1 /2 +(d+1)(d-2)G_1 /d]G_1}{K_1 + 2G_1}} & = \frac{{\phi_2}^2 \mu}{\phi_2 - \sum_{n=2}^\infty D_n}, \label{eq:expansion_shear_modulus}
\end{align}
where $C_n$ and $D_n$ are functionals involving the $n$-point probability functions and double gradients of the appropriate Green's functions, and 
\begin{align}
\kappa &= \frac{K_2 - K_1}{K_2 + 2(d-1)G_1 / d}, \label{eq:bulk_modulus_polarizability}
\\
\mu &= \frac{G_2 - G_1}{ G_2 + \left[dK_1/2 +(d+1)(d-2) G_1/d    \right]G_1/(K_1 +2 G_1)},\label{eq:shear_modulus_polarizability}
\end{align}
are the polarizabilities (strong-contrast parameters) for bulk and shear moduli, respectively.
This expansion is the dynamic analog of the static strong-contrast expansion derived by Torquato \cite{Torquato1997}, which can be viewed as one that perturbs around the Hashin-Shtrikman structures \cite{Hashin1962} that attain the best possible bounds on the effective elastic moduli of isotropic composites for prescribed phase properties and volume fractions.
    In the family of such structures, domains of one phase are topologically disconnected, well-separated from one another, and dispersed throughout a connected (continuous) matrix phase \cite{HashinShtrikman1963, Francfort1986}. 
    This means that strong-contrast expansions will converge rapidly, even for high contrast ratios in phase moduli (Sec. VII.A in the {\it\color{blue}SI}), for dispersions that meet the aforementioned conditions \cite{Rechtsman2008}, and hence the resulting property estimates will be nearly optimal.

\subsection*{Derivation of Eqs. \ref{eq:two-point_Ke}-\ref{eq:two-point_Ge}}
The original strong-contrast approximations [formally identical to Eqs. \ref{eq:two-point_Ke}-\ref{eq:two-point_Ge}] depend on the following two-point parameters $\fn{C_2}{\kL}$ and $\fn{D_2}{\kL}$ are:
\begin{align}
\fn{C_2}{\kL}& = \frac{\pi}{2^{d/2} \fn{\Gamma}{d/2}} \fn{\mathcal{F}}{\kL}\kappa, \label{eq:C2_original}\\
\fn{D_2}{\kL}& = \frac{\pi}{2^{d/2} \fn{\Gamma}{d/2}} \frac{d\cL^2 \fn{\mathcal{F}}{\kT} + 2\cT^2 \fn{\mathcal{F}}{\kL}}{d\cL^2 + 2\cT^2}\mu,\label{eq:D2_original}
\end{align}
where $\fn{\mathcal{F}}{Q}$ is defined in Eq. \ref{eq:A2_original}.
In order to obtain better estimates of $\fn{K_e}{\kL}$ and $\fn{G_e}{\kL}$ in the intermediate-wavelength regime, we modify the strong-contrast approximations in the same manner as we did for the electromagnetic problem.
Specifically, we replace $\fn{\mathcal{F}}{Q}$ in Eqs. \ref{eq:C2_original} and \ref{eq:D2_original} with $\fn{F}{Q}$, defined by Eq. \ref{eq:F_function_Fourier}, which leads to Eqs. \ref{eq:C2} and \ref{eq:D2}. 
The justification for such replacements is based on two observations: (a) $\fn{\mathcal{F}}{Q}$ involves the Green's function of the Helmholtz equation at the wavenumber $Q$, and (b) the associated incident wave should have the wavenumber $Q$.
We note that the modified formulas (Eqs. \ref{eq:two-point_Ke} and \ref{eq:two-point_Ge}) show excellent agreement with numerical simulations; see Sec. V in the {\it\color{blue}SI}.

\subsection*{Spectral density}\label{sec:spectral_density}

For dispersions of nonoverlapping identical spheres of radius $a$, the spectral density can be expressed as \cite{Torquato1985, Torquato_RHM} 
\begin{equation}\label{eq:chik_mono_packing}
\spD{\vect{Q}} = \phi_2 \fn{\tilde{\alpha}_2}{Q;a} \fn{S}{\vect{Q}}, 
\end{equation}
where $\phi_2$ is the packing fraction, $Q\equiv \abs{\vect{Q}}$, $\fn{\tilde{\alpha}_2}{Q;a} \equiv 2^d \pi^{d/2} \fn{\Gamma}{1+d/2} [\fn{J_{d/2}}{Qa}]^2 / Q^d$, and $\fn{J_{\nu}}{x}$ is the Bessel function of the first kind of order $\nu$.
Here, $\fn{S}{\vect{Q}}$ is the structure factor for particle centers, which can be computed from
$$\fn{S}{\vect{Q}} = N^{-1}  \abs{\sum_{j=1}^N \fn{\exp}{-i\vect{Q}\cdot\vect{r}_j}}^2.$$ 
	Therefore, one can easily obtain stealthy hyperuniform and stealthy nonhyperuniform dispersions from the associated point patterns by decorating their point centers with nonoverlapping equal-sized spheres.
	For more details, see Sec. I in the {\it\color{blue}SI}.

\subsection*{Stealthy hyperuniform/nonhyperuniform hard spheres} \label{sec:howtomakestealthy}

We first generate stealthy point configurations in periodic simulation boxes via the collective-coordinate optimization technique \cite{Uche2004,Batten2008,Zhang2015}, i.e., numerical procedures that obtain ground-state configurations for the following potential energy; 
\begin{equation}\label{eq:CC_potential}
\fn{\Phi}{\vect{r}^N} =\frac{1}{v_\mathcal{F}} \sum_{\vect{Q}} \fn{\tilde{v}}{\vect{Q}}\fn{S}{\vect{Q}} +  \sum_{i <j} \fn{u}{r_{ij}},
\end{equation}
where
$$
\fn{\tilde{v}}{\vect{Q}}
=
\begin{cases}
1, & Q_\text{lower}< \abs{\vect{Q}} \leq Q_\text{upper} \\
0, &\mathrm{otherwise}
\end{cases},
$$
 and a soft-core repulsive term \cite{Zhang2017} 
$$
\fn{u}{r}
=
\begin{cases}
(1-r/\sigma)^2, & r < \sigma\\
0,&\mathrm{otherwise}
\end{cases}.
$$
These ground-state configurations are stealthy [i.e., $\fn{S}{\vect{Q}} = 0$ for $Q_\text{lower} < \abs{\vect{Q}} < Q_\text{upper}$] and due to the soft-core repulsions $\fn{u}{r}$, the interparticle distances are larger than $\sigma$.
The resulting point configurations are stealthy hyperuniform if $Q_\text{lower}=0$ but otherwise stealthy nonhyperuniform.
We then circumscribe the points by identical nonoverlapping spheres of radius $a<\sigma/2$.
The reader is referred to Sec. III in the {\it\color{blue}SI} for details. 

\subsection*{Data Availability}  There is no data associated with the manuscript.
}

\showmatmethods{} 

\acknow{The authors gratefully acknowledge the support of the Air Force Office of Scientific Research Program on Mechanics of Multifunctional Materials and Microsystems under award
No. FA9550-18-1-0514.}
\showacknow{}

% Bibliography
%\bibliography{../../ref_elasticwaves}

\end{document}